# Can GLAST detect gamma-rays from the extended radio features of radio galaxies?

R. M. Sambruna*, M. Georganopoulos†, D. Davis**,‡ and A. N. Cillis*

*NASA/GSFC, Code 661, Greenbelt, MD 20771
†Department of Physics, Joint Center for Astrophysics, University of Maryland, Baltimore County, 1000 Hilltop Circle, Baltimore, MD 21250
**CRESST and Astroparticle Physics Laboratory NASA/GSFC, Greenbelt, MD 20771
‡Department of Physics, University of Maryland, Baltimore County, 1000 Hilltop Circle, Baltimore, MD 21250

**Abstract.**
A few FRI radio galaxies were detected at GeV gamma-rays with CGRO EGRET, with peroperties suggesting that the gamma-ray flux originates from the core. Here we discuss the possibility that the extended radio features of radio galaxies could be detected with the LAT, focusing on the particularly promising case of the nearby giant radio galaxy Fornax A.

## GAMMA-RAYS FROM RADIO GALAXIES

Previous observations with *CGRO* EGRET established that the dominant extragalactic sources of GeV gamma-rays are blazars, providing model-independent evidence for beaming in these sources [8]. However, gamma-rays were also detected around the location of the FRI radio galaxies Centaurus A [11] and NGC 6251 [9]. If the GeV flux comes, indeed, from these sources, a most obvious interpretation is that it originates from the compact core of the radio sources. In fact, analysis of the Spectral Energy Distributions (SEDs) from radio to gamma-rays shows a double-humped structure, similar to blazars, but significantly de-beamed [6].

Gamma-ray flux was also detected from the core of more powerful FRIIs, specifically, the Broad-Line Radio Galaxy 3C 111 [10]. Again, the core SED is double-dumped resembling a misaligned blazar.

There are theroetical reasons to expect the cores of radio galaxies to be GeV emitters [10, 5]. Indeed, rescaling for the radio-to-GeV flux ratio of Centaurus A leads to the prediction of significant GeV emission from the cores of roughly two dozen radio sources [6], which *GLAST* should be able to detect in the first 6–18 months of operations.

## EXTENDED RADIO LOBES: THE CASE OF FORNAX A

Can the extended features of radio galaxies be a source of GeV gamma-rays? If so, can they be detected with the LAT? Our analysis of the nearby giant source Fornax A provides a resounding "Yes" to both questions.

The optimal candidates for a *GLAST* LAT detection of the radio lobes should satisfy the following observational criteria:

- Large angular size ($> 30'$) to be resolved with the LAT PSF;
- A weak core, to avoid contamination of the total GeV flux;
- Being located at high Galactic latitudes, to avoid contamination from the unresolved galactic sources.

Finally, good characterization of the radio spectrum is desirable, to constrain the particle energy distribution. Previous detection at X-rays would also be a plus.

One source that fits all these criteria is the nearby (D=18 Mpc) radio galaxy Fornax A, hosted by a massive elliptical. The core of this source is a LINER and its angular size from the E to the W lobe is $33'$. Figure 1a shows the *VLA* image of Fornax A. The lobes were detected with WMAP [1] showing a cutoff of the radio flux above 10 MHz.

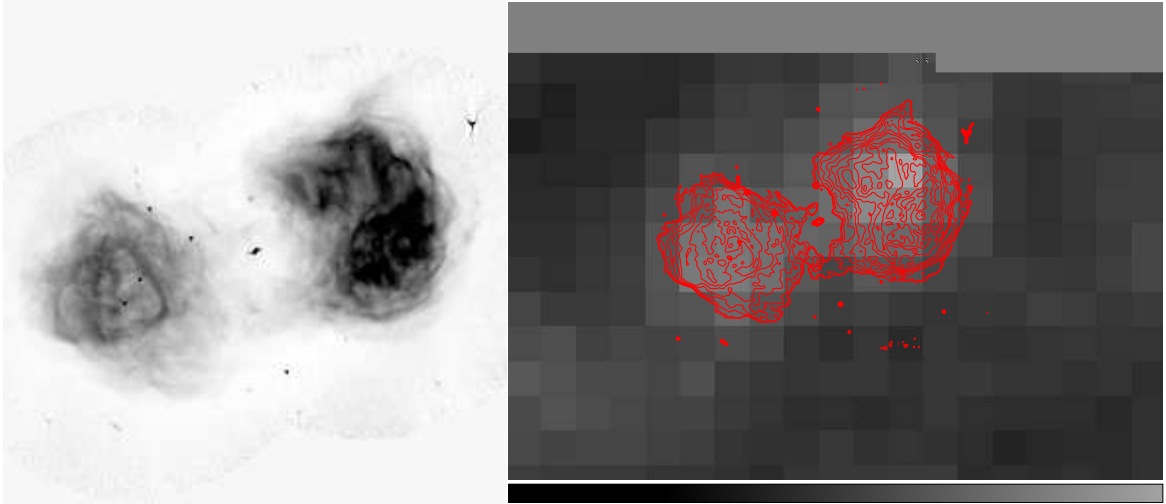

**FIGURE 1.** *(a, Left):* The *VLA* 5 GHz image of Fornax A [4], showing the bright E and W lobes and a weak core. *(b, Right):* Simulation of the LAT image of Fornax A during the first year of the survey (see text). North is up and East to the right.

Interestingly, X-ray emission from both the lobes was previously detected with *ROSAT* [3], *ASCA* [12], and more recently with *XMM-Newton* [7]. The brighter E lobe has a powerlaw X-ray spectrum with slope $\alpha_X \sim 0.6$, consistent with the radio slope below the WMAP cutoff. This supports the idea that the X-rays are produced by Inverse Compton scattering of the CMB photons (IC/CMB) off low-energy ($\gamma \sim 10^3$) electrons in the lobes [3]. The inferred magnetic field is 12$\mu$Gauss [7]. The W lobe is also detected with 1/3 of the X-ray flux.

Can the lobes of Fornax A be detected by the LAT? Assuming the above magnetic field and electron energies, if the electron energy distribution is unbroken (contrary to the WMAP results), the predicted flux for the E lobe is $4 \times 10^{-8}$ ph/cm2/s above 100 MeV. Interestingly, this is the flux measured by EGRET at 2.2$\sigma$ from a reanalysis of the 3rd EGRET catalog [2]. However, if we take into account the WMAP break at 10 GHz, the predicted break energy at gamma-rays is 5 MeV ($1.2 \times 10^{21}$ Hz). If we assume that the spectrum breaks by $\Delta\alpha$=0.5 (from $\alpha_r$=0.62 to $\alpha_r$=1.12), the predicted gamma-ray flux is $1.5 \times 10^{-9}$ ph/cm2/s above 100 MeV.

Using the latter predicted flux and a similar break in the gamma-ray spectrum above 5 MeV, we simulated LAT observations of Fornax A. We assumed that both lobes contribute to the EGRET emission, with the E lobe containing 2/3 of the flux. In 12 months we can detect and resolve gamma-ray emission from the E and W lobe (Fig. 1b), with a total of 28 and 7 counts, respectively.

## CONCLUSIONS

It seems clear, from EGRET observations, that the cores of radio galaxies are gamma-ray sources, easily detectable with the LAT. Using the nearby giant radio galaxy Fornax A we demonstrated that another contender for the origin of gamma-rays from radio sources are the diffuse radio lobes. A simulation shows that both the W and E radio lobes are detected with the LAT in 12 months. Detection of the GeV flux will constrain the electron energy distribution and the magnetic field.

How common is gamma-ray emission from the radio lobes? This issue can be addressed by systematic studies of lobe-dominated radio galaxies with *GLAST*, such as e.g., the 3CRR sample. Feasibility studies are underway.

## ACKNOWLEDGMENTS

We thank the organizers for putting together an interesting and informative meeting. RMS, DD, and AC are supported by the *GLAST* project at GSFC.